\documentclass[aps,prd,twocolumn,superscriptaddress,tightenlines,nofootinbib]{revtex4}

\usepackage{amsfonts,amsmath,amsthm,amssymb}
\usepackage{mathrsfs}
\usepackage{bm}
\usepackage{color}
\usepackage{epsfig}

\usepackage{enumitem}

\newcommand{\PRE}[1]{{#1}}   

\newcommand{\postscript}[2]{\setlength{\epsfxsize}{#2\hsize}
   \centerline{\epsfbox{#1}}}
\newcommand{\comment}[1]{}

\usepackage[usenames,dvipsnames]{xcolor}
\definecolor{orange}{cmyk}{0,0.5,1,0}
\definecolor{rossoCP3}{cmyk}{0,.88,.77,.40}
\definecolor{graa}{rgb}{0.8,0.8,0.8}
\definecolor{blaa}{rgb}{0.2,0.2,0.6}

\begin{document}

\title{\PRE{\vspace*{-0.5in}} \color{rossoCP3}{Neutrinos as a probe of the Universe}}


\author{Luis A. Anchordoqui}
\email{luis.anchordoqui@lehman.cuny.edu}

\affiliation{Department of Physics and Astronomy,  Lehman College, City University of
  New York, NY 10468, USA}

\author{Thomas J. Weiler} \email{tom.weiler@vanderbilt.edu}

\affiliation{Department of Physics and Astronomy, Vanderbilt University, Nashville TN 37235, USA}

\maketitle

The Standard Model (SM) of particle physics is a cornerstone of 20th century physics~\cite{Weinberg:2004kv,Weinberg:2018apv}.  It encompasses the first unification of forces in 90 years, since Maxwell unified electricity and magnetism in the early 1880s by noting that the displacement current unified his equation and made magnetism nothing but moving electricity.

The group theory of SM unification can be written as $SU(3) \times SU(2)_L \times U(1)$, the ``$S$'' indicating that there are $N^2-1$ force particles, one removed from the special (${\rm Det} = +1$) unitary group $U(N)$, while the matter particles of fermions (half-integral quantum mechanical spin) sit in irreducible representations of these three groups~\cite{Glashow:1961tr,GellMann:1961ky,Weinberg:1967tq,Salam:1968rm,Fritzsch:1973pi}.  Not included is gravity, which cannot be a gauge theory as it is not renormalizable.  Gravity/general relativity is a geometric theory~\cite{Einstein:1916vd}.  Thus, we have 8 gluons, 3 $W$ bosons, and a singlet ``photon.''  Adding a complex doublet of Higgs particles to break the $SU(2)_L \times U(1)$ symmetry completes the SM~\cite{Higgs:1964pj}.  The $SU(3)$ is called QCD, short for quantum chromodynamics, and the $SU(2)_L \times U(1)$ when suitably broken will contain quantum electrodynamics (or QED) and weak force.

The SM is a renormalizable field theory, which means that the ultraviolet divergences that appear in loop corrections from perturbation theory can be eliminated by a suitable redefinition of the parameters already appearing in the bare Lagrangian: masses, couplings, and field normalizations. The beta function encodes the dependence of a coupling parameter on the energy scale and so the sign of the beta function determines whether an interaction is asymptotically free or enslaved~\cite{Politzer:1973fx,Gross:1973id}.  The beta function depends on the number of vector force particles and the number of matter particles.  Turns out that only the $SU(3)$ has a Fermi particle count to yield a positive beta function, indicating that $SU(3)= {\rm QCD}$ is confining, while unbroken QED and the weak interaction are not.  The strength of the weak interaction rates are given by the ratio $(m_N/M_W)^2 \sim 10^{-4}$, where $m_N$ is the mass of the nucleon (proton or neutron).  In amplitude, the weak interaction extends to $\sim 1/M_W$, justifying the name weak interaction.  Unification is complete, with the photon arising as the symmetric singlet in the $SU(2)_L \times U(1)$ interaction.  Three higgses appear as the massive modes of the three massive longitudinal $W$ bosons.  There is one higgs left over.  Weak theory and the $U(1)$  have unified as electroweak theory.

\begin{figure}
  \postscript{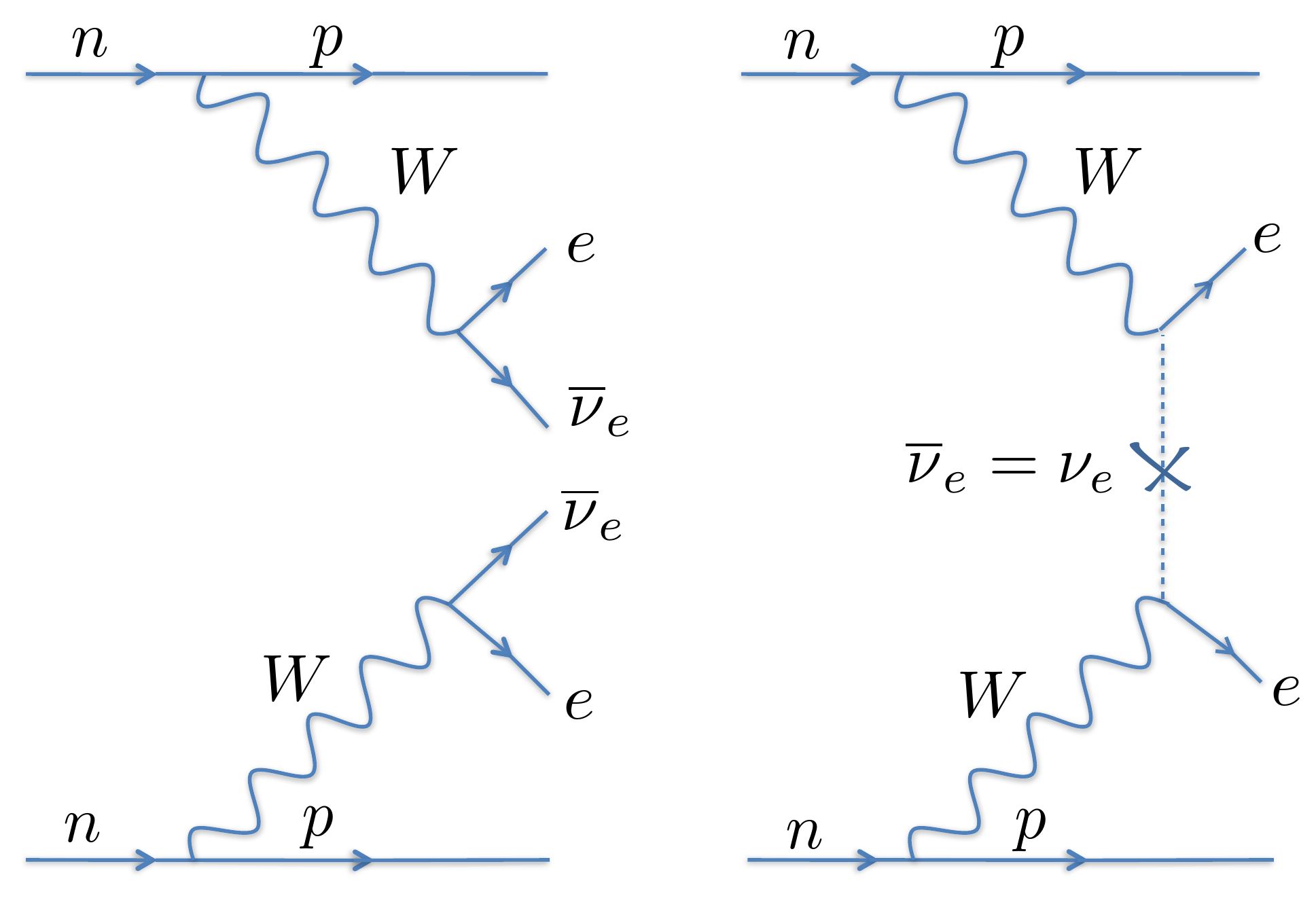}{0.9}
  \caption{The most practical way to uncover the Majorana nature of neutrinos is through the  double beta radioactive decay, in which two  neutrons simultaneously decay into two protons, releasing two electrons and two electron antineutrinos (left panel). This process has been observed in several different elemental isotopes. However, if neutrinos are Majorana particles, the two released antineutrinos can  annihilate each other virtually, yielding  ``neutrinoless double beta decay'' (right panel). Since there are no antineutrinos to carry away any energy, the summed energies of the two electrons should equal the well-known total energy released in the decay. Experiments attempt to observe the antineutrino annihilation by detecting and summing up the energies of the two released electrons. The summed energy of the two electrons normally have a continuous energy spectrum, because some energy is carried away by the antineutrinos. However, if neutrinos are indeed Majorana, there is also a rare chance for the summed energy of the electrons to equal the total energy loss in the decay. It is this small monoenergetic peak that could become the smoking gun for Majorana particles. The rate is proportional to $\sum_j m_{ej} U_{ej}$  and so only expected to be measurable if $j =3$ contains a significant amount of electron neutrino, as in the inverted hierarchy. See main text for details. \label{figura1}}
\end{figure}

Where are the neutrinos?  They are in the matter fermions of the $SU(2)_L$ chiral symmetry.  The symmetry is termed ``chiral'' because the force field, here massless $W$ bosons, couples to left-handed $L$ fermions but not to right-handed $R$ fermions.  $L$ and $R$ are coupled by the mass term.  So, for example, we have the electron with its four states to accommodate two spin states and the two associated antifermion states, but the neutrino, being charge-less, could be its own antiparticle with just two states to accommodate handedness.  Does the neutrino consist of two states or four states?  I.e., is the neutrino like the electron or is it a Majorana fermion~\cite{Majorana:1937vz}, eponymously named after the 1920s physicist who first proposed the self-annihilation of the neutrino?  This is an active area of present investigation.  One ramification, if the neutrino is Majorana, is neutrinoless double beta decay of two nucleons. A Feynman diagram of this process is shown in Fig.~\ref{figura1}.  

Sidney Coleman has said that the symmetry of  vacuum is the symmetry of the world.  In Quantum Field Theory, one may have a vacuum devoid of particles but still containing meaning.  So we need to investigate the symmetry of the vacuum. The vacuum has meaning because of Hawking radiation, which in turn is based upon the Heisenberg uncertainty principle.  It turns out that energy need not be conserved in physics for tiny times, $\Delta E \le \hbar /\Delta t$, where $\hbar$ is a fundamental constant of nature (over a $2 \pi$) named after Planck. The physical picture is this: for small times, a fermion pair, such as electron and positron, may pop out of the vacuum and maintain vacuum quantum numbers as long as it subsequently pops back into the vacuum.
The particles are said to be virtual in that they are not observed.  In fact, neutrinos are never observed, and therefore always virtual.

\begin{figure}
  \postscript{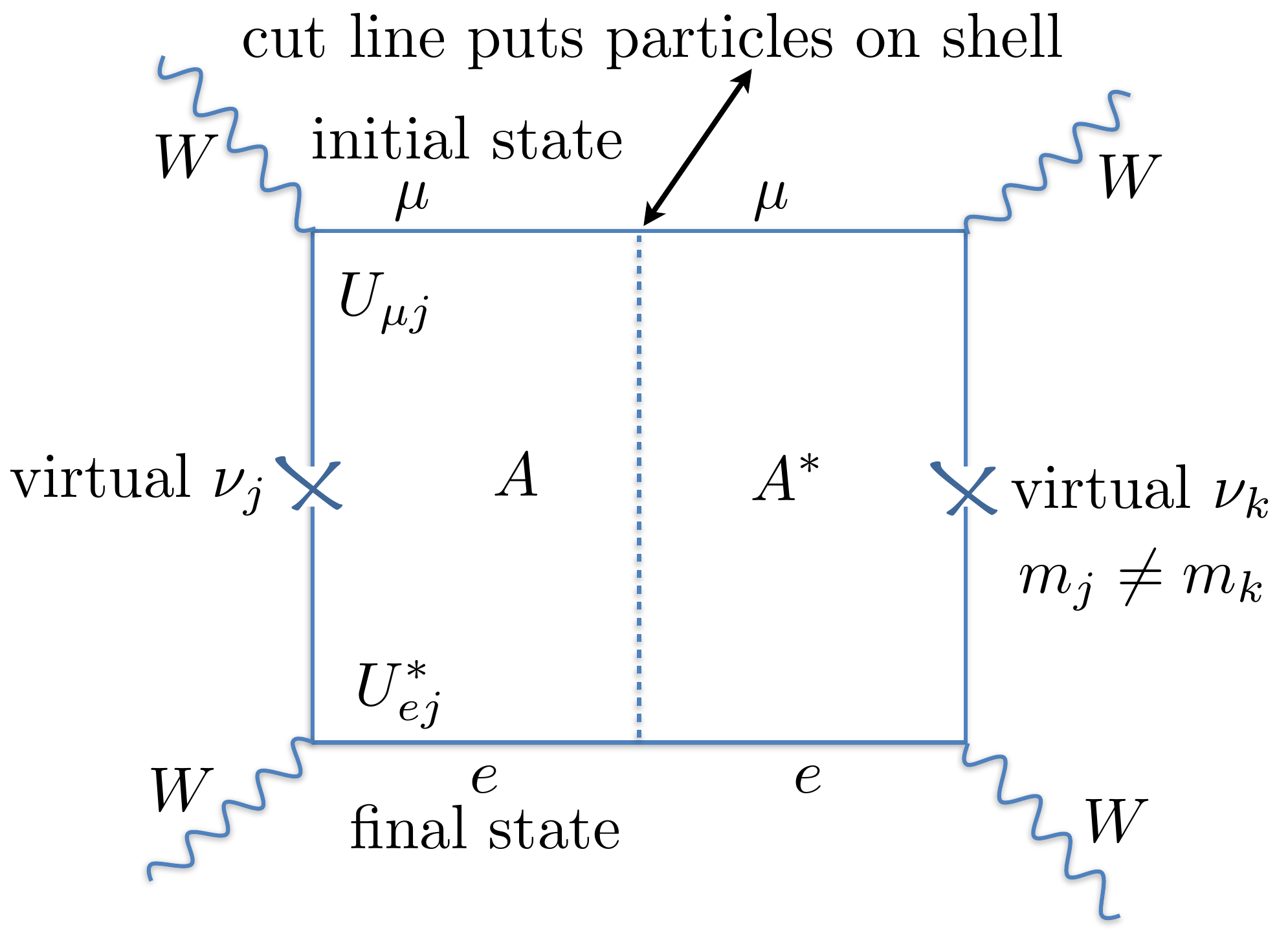}{0.9}
\caption{Cutcosky diagram for neutrino oscillation. \label{figura2}}
\end{figure}

The Nobel prize was awarded a few years back to Takaaki Kajita, director of the SuperK  experiment which discovered oscillations in the late 80s and indicated that neutrinos have a tiny mass, and initiated measurement of Solar neutrinos  based on direction of incidence; and Arthur McDonald, director of SNO lab, for the  measurement of the electron neutrinos  in the Sun which solved the Solar issue~\cite{Fukuda:1998mi,Ahmad:2001an,Ahmad:2002jz}.  Both  experiments revealed the unusual behavior of these misfit particles. Figure~\ref{figura2} shows a Cutcosky diagram for neutrino oscillation.  The dashed line indicates the cut  to of a real axis (on-shell initial and final states).   A neutrino of type $\nu_\mu$ is produced at the $W$ charged current vertex in association a muon $\mu$. The $\nu_\mu$ is virtual. All particles propagate forward in time in mass eigenstates, as shown here for the virtual neutrino. At the lower missing vertex, the neutrino is detected by a second charged current event producing an electron and missing energy. Flavor oscillation has occurred, from an initial $\nu_\mu$ to a final state $\nu_e$.  Left of the cut is the quantum amplitude $A$; right of the cut is $ A^*$.  Probability of occurrence is given by $A A^*=| A|^2$.  Note that quantum mechanics is a theory of amplitudes, not of rates.

The words of a colleague, the neutrino is as 
close to nothing as something can be.  It is charge-less.  Among known interactions, it has only a weak interaction. Formerly believed to be massless, its mass is truly tiny.  Quantum mechanics allows one to pick the basis (with some being more useful than others) to describe the survival probability for flavor transitions.   We have \mbox{$\nu_j= \sum_\alpha U_{j \alpha} \nu_\alpha$,} where $j$ labels the mass states, $\alpha =e,\mu,\tau$ are the three neutrino flavors, and $\mathbb U$ is a large matrix (with elements $U_{j \alpha}$) as demonstrated by experiments~\cite{GonzalezGarcia:2007ib}.  ``Large'' here means the off-diagonal elements are large in appropriate units.  No one knows why elements are large, but we got lucky.  The neutrino quantum has provided many surprises!  Mixing, as with Cabbibo mixing of down and strange quarks, occurs in the Lagrangian. Large mixing in the neutrino sector is a topic under investigation.

Oscillations in space, energy, and time are different from mixing.  They are unique to neutrinos, and evolve only when masses are degenerate enough to fit inside a single quantum mechanical wave function.  They express an interference among mass states which evolves dynamically, meaning in  time/energy/position.  
Oscillation occurs because the three mass state components of the produced flavor travel at slightly different speeds, so that their quantum mechanical wave packets develop relative phase shifts that change how they combine to produce a varying superposition of three flavors. Each flavor component thereby oscillates as the neutrino travels, with the flavors varying in relative strengths. The relative flavor proportions when the neutrino interacts represent the relative probabilities for that flavor of interaction to produce the corresponding flavor of charged lepton. Maki, Nakagawa, and Sakata have chosen the three parameters as angles plus one purely imaginary (the so-called ``CP phase'') reflecting the fact that $\mathbb U= \mathbb U^\dag$ is unitary in quantum mechanics rather than purely real as it is with Euler angles~\cite{Maki:1962mu}. Remember, it is the complex amplitude that is calculated in quantum mechanics, with the real probability then constructed as $P=A A^*=|A|^2$. Experiment has shown that three neutrinos are needed for complexity to appear in neutrino physics~\cite{Barger:1998ta}.  And yet we know that oscillations divide into a basis of two $2 \times 2$ sums, with a ratio of about 30.  This is one reason why the complex CP phase is so hard to deduce.  From solar neutrino data we furthermore know that $\nu_2$ couples with the lighter sum.  The question arises: is $\nu_3$ heavier than $\nu_2$ or lighter?  This question is also under investigation.  Also under investigation are the mixing angles, e.g. is the largest mixing angle larger or smaller than maximal mixing at $45^\circ$? 

Another gift is offered to us by the small neutrino masses, namely, the generation of $\nu$ masses.   Fine tuning of the neutrino's Yukawa couplings, as is done with all other particles, is a possibility. But the neutrino is known to be five or more orders of magnitude lighter than the next lightest particle, the electron, so an independent mechanism has been invented, the ``see-saw mechanism.'' The see-saw uses properties of a simple $2 \times 2$ matrix.  The matrix lives in neutrino space, with a zero naturally place in the $\nu$-$\nu$ $2 \times 2=3+1$ (group theory numbers) space, a weak coupling between $\nu$ and $N$  $(2+0)$, and a heavy $N (0+0)$.  Diagonalization then yields the correct experimental neutrino mass M $\sim(M^2_W/M_N)$ $\sim~{\rm eV}$ for $M_N \sim 10^{23}~{\rm eV}$~\cite{Mohapatra:1979ia}.  

In both the see-saw and the Yukawa coupling models, all particle masses are generated by the interactions of the particle's field with the vacuum state.  Particles are slowed by vacuum interactions.  According to special relativity, slowness is in turn related to the perceived mass.  Slowness rises with mass.  So massless particles have the least slowness.  This does not mean that massless particles, like the photon and probably the graviton, are impervious to the vacuum.  E.g. the photon's vacuum numbers are not zero, but rather $c^2 =1/(\epsilon_0 \mu_0)$.

The vacuum also provides a measurable difference between classical and quantum mechanics~\cite{Einstein:1935rr}.  John Bell found that difference~\cite{Bell:1964kc}. In articulating his famous inequality 50 years ago, he said ``For me, it is so reasonable to assume that the photons in those experiments carry with them programs, which have been correlated in advance, telling them how to behave. This is so rational that I think that when Einstein saw that, and the others refused to see it, he was the rational man. ... for me, it is a pity that Einstein's idea doesn't work. The reasonable thing just doesn't work.''  It is clear that  Bell thought and hoped hidden variables would be the answer.  But they were not the answer, and the present interpretation is that quantum mechanics requires an observer to put particles on their mass shells.

Neutrinos' low mass and neutral charge mean they interact exceedingly weakly with other particles and fields. This feature of weak interaction is of interest particularly because it means neutrinos can be used to probe environments that other radiation (such as light or radio waves) cannot penetrate.  
Using neutrinos as a probe was first proposed in the mid-20th century as a way to detect conditions at the core of the Sun. The solar core cannot be imaged directly because electromagnetic radiation (such as light) is diffused by the great amount and density of matter surrounding the core. On the other hand, neutrinos pass through the Sun with few interactions. Whereas photons emitted from the solar core may require 40,000 years to diffuse to the outer layers of the Sun, neutrinos generated in stellar fusion reactions at the core cross this distance practically unimpeded at nearly the speed of light. 

Solar neutrinos originate from the nuclear fusion powering the Sun and other stars. The details of the operation of the Sun are explained by Bahcall's standard solar model~\cite{Bahcall:2002ng}. In short: when four protons fuse to become one helium nucleus, two of them have to convert into neutrons, and each such conversion releases one electron neutrino.  The Sun sends enormous numbers of neutrinos in all directions. Each second, about 65 billion (i.e. $6.5 \times 10^{10}$) solar neutrinos pass through every square centimeter on the part of the Earth orthogonal to the direction of the Sun.  Since neutrinos are insignificantly absorbed by the mass of the Earth, the surface area on the side of the Earth opposite the Sun receives about the same number of neutrinos as the side facing the Sun.

Neutrinos can play an important role in the evolution of the universe, modifying some of the cosmological observables~\cite{Lee:1977ua}. It is thought that, just like the cosmic microwave background radiation left over from the Big Bang, there is a background of low-energy neutrinos in our Universe. In the 1980s it was proposed that these may be the explanation for the dark matter thought to exist in the universe. Neutrinos have one important advantage over most other dark matter candidates: They are known to exist. This idea also has serious problems. From particle experiments, it is known that neutrinos are very light. This means that they easily move at speeds close to the speed of light. For this reason, dark matter made from neutrinos is termed ``hot dark matter.'' The problem is that being fast moving, the neutrinos would tend to have spread out evenly in the universe before cosmological expansion made them cold enough to congregate in clumps. This would cause the part of dark matter made of neutrinos to be smeared out and unable to cause the large galactic structures that we see. 
From cosmological measurements, it has been calculated that the sum of the three neutrino masses must be less than one millionth that of the electron~\cite{Aghanim:2018eyx}. The relic background neutrinos are estimated to have a density of $56$ of each type per cubic centimeter and temperature $1.9~{\rm K}$ (or $1.7 \times 10^{-4}~{\rm eV}$) if they are massless, much colder if their mass exceeds $0.001~{\rm eV}$. Although their density is quite high, relic neutrinos have not yet been observed in the laboratory, as their energy is below thresholds of most detection methods, and due to extremely low neutrino interaction cross-sections at sub-eV energies. In contrast, boron-8 solar neutrinos --which are emitted with a higher energy-- have been detected definitively despite having a space density that is lower than that of relic neutrinos by some 6 orders of magnitude.

In 1966, Colgate and White calculated that neutrinos carry away most of the gravitational energy released by the collapse of massive stars, events now categorized as Type Ib and Ic and Type II supernovae~\cite{Colgate:1966ax}. When such stars collapse, matter densities at the core become so high ($10^{17}~{\rm kg/m^3}$) that the degeneracy of electrons is not enough to prevent protons and electrons from combining to form a neutron and an electron neutrino. A second and more profuse neutrino source is the thermal energy (100 billion kelvins) of the newly formed neutron core, which is dissipated via the formation of neutrino-antineutrino pairs of all flavors. The core collapse phase of a supernova is an extremely dense and energetic event. It is so dense that no known particles are able to escape the advancing core front except for neutrinos. Consequently, supernovae are known to release approximately 99\% of their radiant energy in a short (10 second) burst of neutrinos. These neutrinos are a very useful probe for core collapse studies. For an average supernova, approximately $10^{57}$ (an octodecillion) neutrinos are released, but the actual number detected at a terrestrial detector will be far smaller; e.g., a dozen neutrinos observed by the Kamiokande II experiment  came from Supernova 1987A~\cite{Hirata:1987hu}, which occurred in the Large Magellanic Cloud, a small galaxy that orbits the Milky Way.

Neutrinos are also useful for probing astrophysical other sources beyond the Solar System because they are the only known particles that are not significantly attenuated by their travel through the interstellar medium~\cite{Gaisser:1994yf,Anchordoqui:2013dnh}. Optical photons can be obscured or diffused by dust, gas, and background radiation. Ultra-high-energy cosmic rays~\cite{Anchordoqui:2018qom}, in the form of swift protons and atomic nuclei, are unable to travel more than about 100 megaparsecs due to the Greisen-Zatsepin-Kuzmin limit (GZK cutoff)~\cite{Greisen:1966jv,Zatsepin:1966jv}. Neutrinos, in contrast, can travel even greater distances barely attenuated. The exception are Zettavolt neutrinos which would annihilate resonantly on the relic-neutrino background to produce a ``Weiler-burst'' of SM particles~\cite{Weiler:1982qy}.

In 1987 Francis Halzen brought forth the means for ``ice-fishing'' cosmic neutrinos. In July 2018, the IceCube neutrino
observatory announced that they have traced an extremely-high-energy neutrino that hit their Antarctica-based research station in September 2017 back to its point of origin in the blazar TXS 0506+056 located 3.7 billion light-years away in the direction of the constellation Orion~\cite{IceCube:2018dnn}. This is the first time that a neutrino detector has been used to locate an object in space and that a source of cosmic rays has been identified. Very recently, IceCube detected~\cite{IceCube:2021rpz} Glashow's resonant scattering of electron antineutrinos on electrons~\cite{Glashow:1960zz}, opening a new window to study astrophysical sources and particle physics~\cite{Anchordoqui:2004eb}. 
Several topics in neutrino physics were not included in this report due to brevity concerns.  These include the Jarlskog invariant for CP violation in the leptonic sector, the possibility of extra space dimensions with membrane-bulk resonances of neutrinos, or even no space itself through spin networks of neutrinos,  the conflict between quantum mechanics and general relativity, black holes and holography, and more on sterile neutrinos as possible dark matter.  Some of these topics are quite speculative, but all point to the opportunities for research provided by the simple neutrino. \\

L.A.A. is supported by the U.S. National Science Foundation (NSF) Grant PHY-1620661 and the National Aeronautics and Space Administration (NASA) Grant 80NSSC18K0464. T.J.W. is supported by the DoE Grant DE-SC-0011981.

\end{document}